\documentclass[a4paper]{article}
\usepackage{amsfonts}
\usepackage{amssymb}
\usepackage[margin=2cm]{geometry}
\usepackage[dvips]{graphicx}

\usepackage[T1]{fontenc}
\usepackage[latin2]{inputenc}

\usepackage{amsmath}
\usepackage{bbm}

\begin{document}


\title{Comments on {\sl Fault-Tolerant Quantum Computation for Local Non-Markovian Noise}}

\author{Robert Alicki\\ 
  Institute of Theoretical Physics and Astrophysics, University
of Gda\'nsk, \\ Wita Stwosza 57, PL 80-952 Gda\'nsk, Poland}

\date{\today}
\maketitle


In the recent paper \cite{Ter} Terhal and Burkard derived a threshold result for fault-tolerant quantum computation under the assumption of the non-Markovian noise and claimed to rebut the objections rised in \cite{AH3}. The purpose of this note is to show that the main condition used in \cite{Ter}, although looking quite innocently, implies the assumption of the extremally low probability error per single quantum gate
$p \leq 10^{-8} - 10^{-12}$
i.e. the square of the expected threshold value for the case of Markovian noise.
 
The basic results of \cite{Ter} involve two parameters: the duration of the elementary gate $t_0$ and the energy scale $\lambda_0$ which provides an upper bound on the single qubit interaction with the bath
\begin{equation}
\|H_{SB}[q_i]\| \leq \lambda_0\ .
\label{2}
\end{equation}
The threshold condition depending on the code is given by
\begin{equation}
\lambda_0 t_0 <  10^{-4} - 10^{-6}\ .
\label{3}
\end{equation}
As explained in \cite{AH3} the frequency dependent non-Markovian decoherence time $\tau_D(\omega)$ is given in terms of the spectral density $R(\omega)$ characterizing the properties of the system-bath interaction and dependent on the bath's dynamics.
Assume the simple form of the single qubit interaction 
\begin{equation}
H_{SB}[q_i]= \sigma \otimes A
\label{4}
\end{equation}
where $\sigma$ represents dimensionless (O(1)-order) qubit operator and $A$ is the bath's observable with the dimension of energy ( we put $\hbar = 1$). For simplicity we can use the following Anzatz for the bath's autocorrelation function
\begin{equation}
<A(t)A> = \lambda^2 \exp(-t/\tau_B)
\label{5}
\end{equation}
where due to (\ref{2}),(\ref{4}) $\lambda \leq \lambda_0$ and $\tau_B$ is the relaxation time for the bath. Hence we obtain a Lorentzian shape of the spectral density being the Fourier transform of the autocorrelation funcion (\ref{5})
\begin{equation}
R(\omega) = \lambda^2 \frac{1/\tau_B}{\omega^2 + (1/\tau_B)^2}\ .
\label{6}
\end{equation}
The frequency dependent decoherence time $\tau_D(\omega)$ satisfies
\begin{equation}
\tau_D(\omega) \simeq  [R(\omega)]^{-1} = \frac{1}{\lambda^2\tau_B}\bigl(1  + (\tau_B\omega)^2\bigr)>\frac{1}{\lambda^2\tau_B}\ .
\label{7}
\end{equation}
and for all reasonable models of reservoirs their
relaxation time $\tau_B$ is the shortest relevant time scale or at most comparable with the time scale of the qubit Hamiltonian
$\sim t_0$. Therefore the decoherence time satisfies 
\begin{equation}
\tau_D \simeq  \frac{1}{\lambda^2\tau_B} > \frac{1}{\lambda_0^2 t_0^2}\frac{t_0}{\tau_B} t_0 \geq 
\Bigl(\frac{1}{\lambda_0^2 t_0^2}\Bigr)t_0 \ .
\label{8}
\end{equation}
Using (\ref{3}) we obtain the following estimation for the error probability per single gate
\begin{equation}
p = \frac{t_0}{\tau_D} < 10^{-8} - 10^{-12}\ .
\label{9}
\end{equation}
Obviously, the final estimation does not depend on the details of the autocorrelation function (\ref{5}) but only the time and energy scales given by $t_0, \tau_B$ and $\lambda_0$ matter.

\end{document}